\documentclass{sf2a-conf}
\usepackage{graphicx}

\begin{document}
\TitreGlobal{SF2A 2007}
\title{The formation of Spiral Arms and Rings in Barred Galaxies }


\author{M. Romero-G\'omez} \address{ LAM, Observatoire Astronomique de Marseille-Provence, 2 Place Le Verrier, 13248 Marseille, France}

\runningtitle{Spiral Arms and Rings in Barred Galaxies}

\setcounter{page}{1}

\index{Romero-G\'omez M.}

\maketitle

\begin{abstract}
We propose a theory to explain the formation of both spirals and rings in 
barred galaxies using a common dynamical framework. It is based on the orbital 
motion driven by the unstable equilibrium points of the rotating bar potential. 
Thus, spirals, rings and pseudo-rings are related to the invariant manifolds 
associated to the periodic orbits around these equilibrium points. We examine 
the parameter space of three barred galaxy models and discuss the formation 
of the different morphological structures according to the properties of the 
bar model. We also study the influence of the shape of the rotation curve in 
the outer parts, by making families of models with rising, flat or falling 
rotation curves in the outer parts. The differences between spiral and ringed 
structures arise from differences in the dynamical parameters of the host 
galaxies.
\end{abstract}

\section{Introduction}
Bars are a very common feature of disc galaxies. In a sample of $186$
spirals drawn from the Ohio State University Bright Spiral Galaxy
Survey, Eskridge {\it et al.} (\cite{esk00}) find that $56\%$ of the
galaxies in the near infrared are strongly barred, while an additional 
$6\%$ are weakly barred. A large fraction of barred galaxies show two 
clearly defined spiral arms (e.g. Elmegreen \& Elmegreen \cite{elm82}), 
often departing from the end of the bar at nearly right angles. This is 
the case for instance in NGC~1300, NGC~1365 and NGC~7552. Deep exposures show 
that these arms wind around the bar structure and extend to large distances 
from the centre (see for instance Sandage \& Bedke \cite{san94}). Almost all 
researchers agree that spiral arms and rings are driven by the gravitational 
field of the galaxy (see Toomre \cite{too77} and Athanassoula \cite{ath84}, 
for reviews). In particular, spirals are believed to be density waves in 
a disc galaxy (Lindblad \cite{lind63}). Toomre (\cite{too69}) found that 
the spiral waves propagate towards the principal Lindblad resonances of the 
galaxy, where they damp down, and thus concludes that long-lived spirals need 
some replenishment. There are essentially three different possibilities for a 
spiral wave to be replenished. First, it can be driven by a companion or 
satellite galaxy. A direct, relatively slow, and close passage of another 
galaxy can form trailing shapes (e.g. Toomre \cite{too69}; Toomre \& Toomre 
\cite{too72}; Goldreich \& Tremaine \cite{gol78}, \cite{gol79}; Toomre 
\cite{too81} and references therein). They can also be excited by the 
presence of a bar. Several studies have shown that a rotating bar or oval can 
drive spirals (e.g. Lindblad \cite{lind60}; Toomre \cite{too69}; Sanders 
\& Huntley \cite{san76}; Schwarz \cite{sch79}, \cite{sch81}; Huntley 
\cite{hun80}). The third alternative, proposed by Toomre (\cite{too81}), is 
the swing amplification feedback cycle. This starts with a leading wave 
propagating from the centre towards corotation. In doing so, it unwinds and 
then winds in the trailing sense, while being very strongly amplified. 
This trailing wave will propagate towards the centre, while a further trailing 
wave is emitted at corotation and propagates outwards, where it is dissipated 
at the Outer Lindblad Resonance. The inwards propagating trailing wave, when 
reaching the centre will reflect into a leading spiral, which will propagate 
outwards towards corotation, thus closing the feedback cycle. 
Danby (\cite{dan65}) argued that orbits in the gravitational potential
of a bar play an important role in the formation of arms. He noted that orbits 
departing from the vicinity of the equilibrium points located at the ends of 
the bar describe loci with the shape of spiral arms and can be responsible for 
the transport of stars from within to outside corotation, and vice versa. 
Unfortunately, he did not set his work in a rigorous theoretical context, so 
that it remained purely phenomenological. He also investigated whether orbits 
can be responsible for ring-like structures, but in this case, he did not 
consider orbits departing from the ends of the bar as he previously did when
accounting for the spiral arms. 

Strongly barred galaxies can also show prominent and spectacular rings or 
partial rings. The origin of such morphologies has been studied by Schwarz 
(\cite{sch81}, \cite{sch84}, \cite{sch85}), who followed the response of a 
gaseous disc galaxy to a bar perturbation. He proposed that ring-like patterns 
are associated to the principal orbital resonances, namely ILR, CR, and OLR. 
There are different types of outer rings. Buta (\cite{but95}) classified them 
according to the relative orientation of the ring and bar major axes. If these 
two axes are perpendicular, the outer ring is classified as $R_1$. If the two 
axes are parallel, the outer ring is classified as $R_2$. Finally, if both 
types of rings are present in the galaxy, the outer ring is classified as $R_1R_2$.

In Romero-G\'omez {\it et al.} (\cite{rom06,rom07}), we propose that rings 
and spiral arms are the result of the orbital motion driven by the invariant 
manifolds associated to periodic orbits around unstable equilibrium points. 
In Romero-G\'omez {\it et al.} (\cite{rom06}), we fix a barred galaxy 
potential and we study the dynamics around the unstable equilibrium points.
We give a detailed definition of the invariant manifold associated to a 
periodic orbit. For the model considered, the invariant manifolds delineate
well the loci of an $rR_1$ ring structure. In Romero-G\'omez {\it et al.}
(\cite{rom07}), we construct families of models based on simple, yet realistic, 
barred galaxy potentials. In each family, we vary one of the free parameters of 
the potential and keep the remaining fixed. For each model, we numerically 
compute the orbital structure associated to the invariant manifolds.
In this way, we are able to study the influence of each model parameter on the
global morphologies delineated by the invariant manifolds.

In Sect. \ref{sec:mod}, we first present the equations of motion and the 
galactic models used in the computations. In Sect. \ref{sec:inv}, we give a 
brief description of the dynamics around the unstable equilibrium points. 
In Sect. \ref{sec:res}, we show the different morphologies that result from 
the computations. In Sect. \ref{sec:dis}, we compare our results with some 
observational features and conclude.

\section{Equations of motion and description of the model}
\label{sec:mod}
We model the potential of a barred galaxy as the superposition of three
components, two of them axisymmetric and the third bar-like. The
last component rotates anti-clockwise with angular velocity ${\bf
\Omega_p}=\Omega_p{\bf z}$, where $\Omega_p$ is a constant pattern
speed\, \footnote{Bold letters denote vector notation. The vector {\bf z} is a unit
vector.}. The equations of motion in 
this potential in a frame rotating with angular speed ${\bf \Omega_p}$ in vector form are
\begin{equation}\label{eq-motvec}
{\bf
\ddot{r}=-\nabla \Phi} -2{\bf (\Omega_p \times \dot{r})-  \Omega_p \times
(\Omega_p\times r)},
\end{equation}
where the terms $-2 {\bf \Omega_p\times \dot{r}}$ and $-{\bf \Omega_p \times
(\Omega_p\times r)}$ represent the Coriolis and the centrifugal
forces, respectively, and ${\bf r}$ is the position vector. We define an 
effective potential $\Phi_{\hbox{\scriptsize eff}}=\Phi-\frac{1}{2}\Omega_p^2\,
(x^2+y^2),$ then Eq. (\ref{eq-motvec}) becomes ${\bf \ddot{r}=-\nabla \Phi_{\hbox{\scriptsize eff}}} -2{\bf (\Omega_p \times \dot{r})},$ and the Jacobi constant is 
\begin{equation}\label{eq-energy}
E_J = \frac{1}{2} {\bf\mid \dot{r}\mid} ^2 + \Phi_{\hbox{\scriptsize eff}},
\end{equation}
which, being constant in time, can be considered as the energy in the
rotating frame.

The axisymmetric component consists of the superposition of a disc 
and a spheroid. The disc is modelled as a Kuzmin-Toomre disc 
(Kuzmin \cite{kuz56}; Toomre \cite{too63}) of surface density $\sigma(r)$ and
the spheroid is modelled using a density distribution of the form $\rho(r)$:
\begin{equation}\label{eq:kuz-sph}
\sigma(r) = \frac{V_d^2}{2\pi r_d}\left(1+\frac{r^2}{r_d^2}\right)^{-3/2}, \qquad
\rho(r)=\rho_b\left(1+\frac{r^2}{r_b^2}\right)^{-3/2}.
\end{equation}
The parameters $V_d$ and $r_d$ set the scales of the velocities and radii of the
disc, respectively, and  $\rho_b$ and $r_b$ determine the concentration and 
scale-length of the spheroid. In our models, we use three bar potentials to compare
the results obtained. The first bar potential is described by a Ferrers 
(\cite{fer77}) ellipsoid whose density distribution is:
\begin{equation}
\left\{\begin{array}{lr}
\rho_0(1-m^2)^n & m\le 1\\
 0 & m\ge 1,
\end{array}\right.
\label{eq:Ferden}
\end{equation}
where $m^2=x^2/a^2+y^2/b^2$. The values of $a$ and $b$ determine the shape of 
the bar, $a$ being the length of the semi-major axis, which is placed along 
the $x$ coordinate axis, and $b$ being the length of the semi-minor axis. The
parameter $n$ measures the degree of concentration of the bar and $\rho_0$ 
represents the bar central density. We also use two ad-hoc potentials, namely 
a Dehnen's bar type, $\Phi_1$, (Dehnen \cite{deh00}) and a Barbanis-Woltjer 
(BW) bar type, $\Phi_2$, (Barbanis \& Woltjer \cite{bar67}):
\begin{equation}
\Phi_1(r,\theta)=-\frac{1}{2}\epsilon v_0^2\cos(2\theta)\left\{{\begin{array}{ll}
\displaystyle 2-\left(\frac{r}{\alpha}\right)^n, & r\le \alpha\rule[-.5cm]{0cm}{1.cm}\\
\displaystyle \left(\frac{\alpha}{r}\right)^n, & r\ge \alpha.\rule[-.5cm]{0cm}{1.cm}
\end{array}}\right. \qquad \Phi_2(r,\theta)=\hat{\epsilon}\sqrt{r}(r_1-r)\cos(2\theta)
\label{eq:adhoc}
\end{equation}
The parameter $\alpha$ is a characteristic length scale of the Dehnen's type bar 
potential, and $v_0$ is a characteristic circular velocity. The parameter 
$\epsilon$ is related to the bar strength. The parameter $r_1$ is a characteristic 
scale length of the BW bar potential and the parameter $\hat{\epsilon}$ is related 
to the bar strength.

\section{Dynamics around the $L_1$ and $L_2$ equilibrium points}
\label{sec:inv}
\begin{figure}[h]  
\centering 
\includegraphics[width=5cm,angle=-90.]{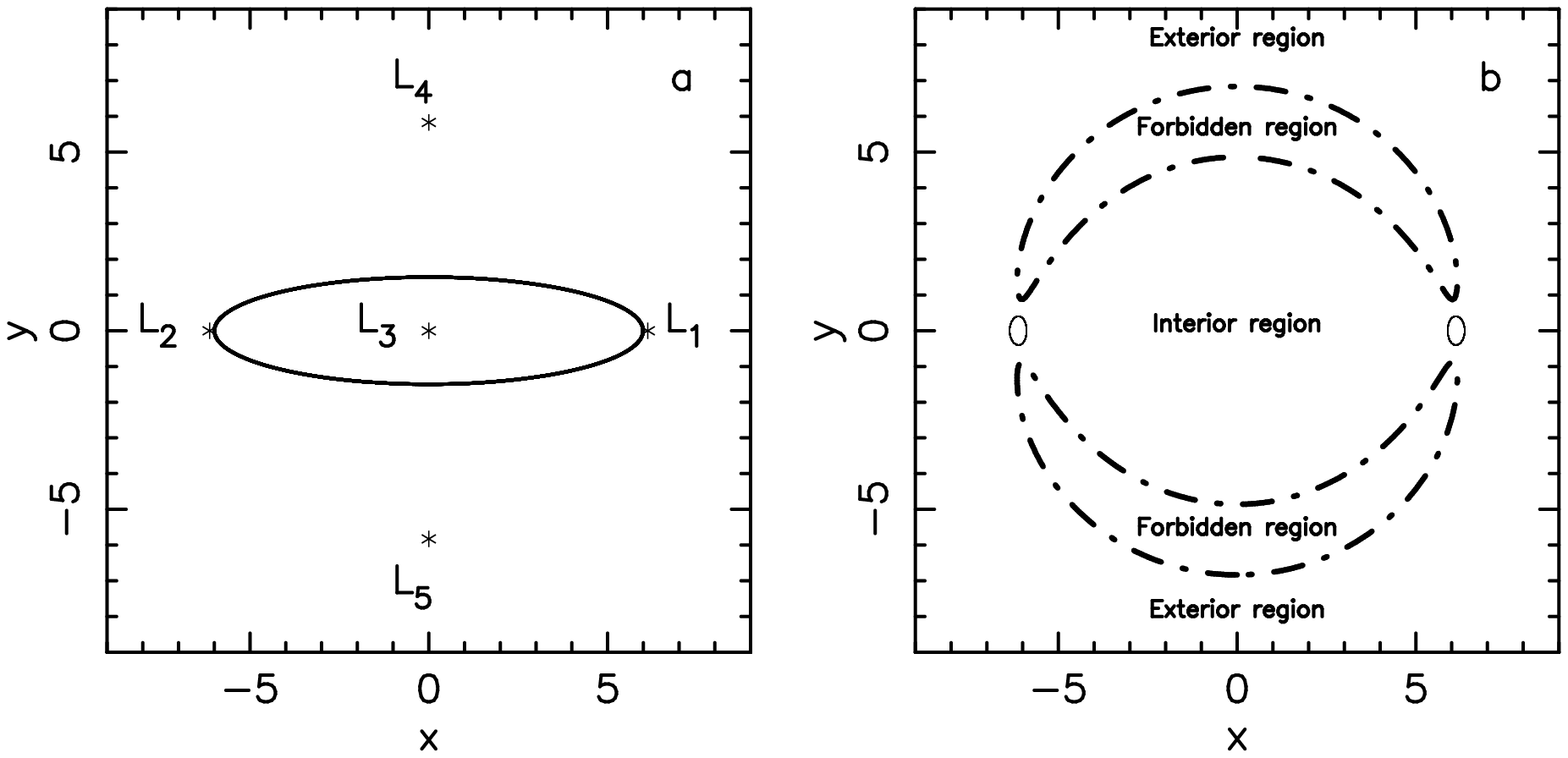}\hspace{0.25cm}
\includegraphics[width=4.65cm,angle=-90.]{romerogomez_png_fig2.ps}
\caption{Dynamics around the $L_1$ and $L_2$ equilibrium points. {\it Left 
panel:} location of the equilibrium points and outline of the bar. 
{\it Middle panel:} Zero velocity curves and Lyapunov orbits around $L_1$
and $L_2$. {\it Right panel:} Stable, $W^s_{\gamma_1}$ in green, and unstable, 
$W^u_{\gamma_1}$ in red, invariant manifolds of a periodic orbit around $L_1$.} 
\label{myfig1} 
\end{figure}

For our calculations we place ourselves in a frame of 
reference corotating with the bar, and the bar semi-major axis is located
along the $x$ axis. In this rotating frame we have five equilibrium
Lagrangian points (see left panel of Fig.~\ref{myfig1}). Three of these points 
are stable, namely $L_3$, which is placed at the centre of the system, and 
$L_4$ and $L_5$, which are located symmetrically on the $y$ axis. $L_1$ and $L_2$ 
are unstable and are located symmetrically on the $x$ axis. The surface 
$\Phi_{\hbox{\scriptsize eff}}=E_J$ ($E_J$ defined as in Eq. (\ref{eq-energy})) 
is called the zero velocity surface, and its intersection with the $z=0$ plane 
gives the zero velocity curve. All regions in which 
$\Phi_{\hbox{\scriptsize eff}}>E_J$ are forbidden to a star with this 
energy, and are thus called forbidden regions. The zero velocity curve also 
defines two different regions, namely, an exterior region and an interior one 
that contains the bar. The interior and exterior regions are connected via the 
equilibrium points (see middle panel of Fig.~\ref{myfig1}). Around the equilibrium 
points there exist families of periodic orbits, e.g. around the central 
equilibrium point the well-known $x_1$ family of periodic orbits that is 
responsible for the bar structure. 

The dynamics around the unstable equilibrium points is described in detail in
Romero-G\'omez {\it et al.} (\cite{rom06}), here we give a brief summary.
Around each unstable equilibrium point there also exists a family of periodic 
orbits, known as the family of Lyapunov orbits (Lyapunov \cite{lya49}). For a 
given energy level, two stable and two unstable sets of asymptotic orbits emanate 
from the periodic orbit, and they are known as the stable and unstable invariant 
manifolds, respectively. We denote by $W^s_{\gamma_i}$ the stable invariant 
manifold associated to the periodic orbit $\gamma$ around the equilibrium 
point $L_i,\, i=1,2$. This stable invariant manifold is the set of 
orbits that tends to the periodic orbit asymptotically. In the same way we 
denote by $W^u_{\gamma_i}$ the unstable invariant manifold associated to the 
periodic orbit $\gamma$ around the equilibrium point $L_i,\, i=1,2$. This unstable 
invariant manifold is the set of orbits that departs asymptotically from the 
periodic orbit (i.e. orbits that tend to the Lyapunov orbits when the time 
tends to minus infinity), see right panel of Fig.~\ref{myfig1}. Since the 
invariant manifolds extend well beyond the neighbourhood of the equilibrium 
points, they can be responsible for global structures. 

In Romero-G\'omez {\it et al.} (\cite{rom07}), we give a detailed description
of the role invariant manifolds play in global structures, in particular, in
the transfer of matter. Simply speaking, the transfer of matter is characterised 
by the presence of homoclinic, heteroclinic, and transit orbits.

\begin{figure}[h]  
\centering 
\includegraphics[width=5.cm,angle=-90.]{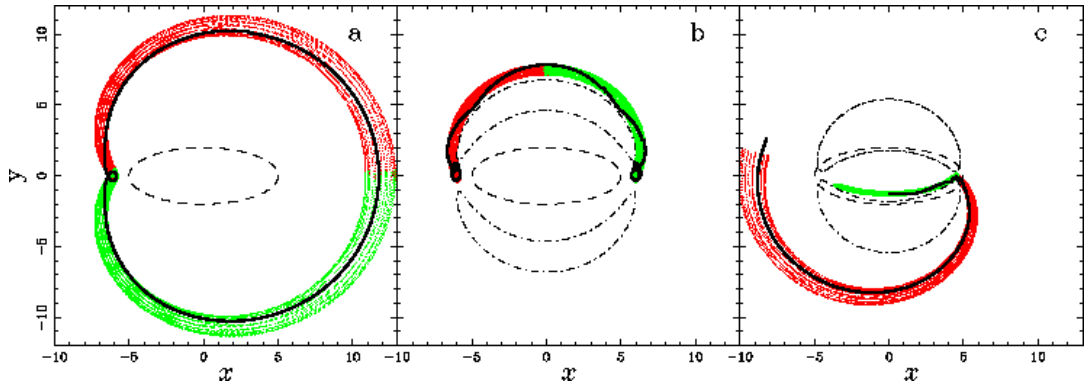}
\caption{Homoclinic {\bf (a)}, heteroclinic {\bf (b)} and escaping
{\bf (c)} orbits (black thick lines) in the configuration space. In
red lines, we plot the unstable invariant manifolds associated to the
periodic orbits, while in green we plot the corresponding stable invariant
manifolds. In dashed lines, we give the outline of the bar and, in 
{\bf (b)} and {\bf (c)}, we plot the zero velocity curves in dot-dashed lines.} 
\label{myfig2}
\end{figure}

Homoclinic orbits correspond to asymptotic trajectories, $\psi$, such that 
$\psi\in W^u_{\gamma_i}\cap W^s_{\gamma_i},\,i=1,2$. Thus, a homoclinic orbit 
departs asymptotically from the unstable Lyapunov periodic orbit $\gamma$ around 
$L_i$ and returns asymptotically to it (see Fig.~\ref{myfig2}a). 
Heteroclinic orbits are asymptotic trajectories, $\psi^\prime$, such that
$\psi^\prime\in W^u_{\gamma_i}\cap W^s_{\gamma_j},\, i\ne j,\,i,j=1,2$. Thus, 
a heteroclinic orbit departs asymptotically from the periodic orbit 
$\gamma$ around $L_i$ and asymptotically approaches the corresponding Lyapunov 
periodic orbit with the same energy around the Lagrangian point at the opposite
end of the bar $L_j$, $i\ne j$ (see Fig.~\ref{myfig2}b). There also exist 
trajectories that spiral out from the region of the unstable periodic orbit, and
we refer to them as transit orbits (see Fig.~\ref{myfig2}c).

\section{Results}
\label{sec:res}
Here we describe the main results obtained when we vary the parameters of the models
introduced in Sect. \ref{sec:mod}. In order to best see the influence of each
parameter separately, we make families of models in which only one of the
free parameters is varied, while the others are kept fixed. Our results in
Romero-G\'omez {\it et al.} (\cite{rom07}) show that only the bar pattern 
speed and the bar strength have an influence on the shape of the invariant 
manifolds, and thus, on the morphology of the galaxy. We have also studied the 
influence of the shape of the rotation curve. We make models with either rising, 
flat or falling rotation curve in the outer parts. 

Our results also show that the morphologies obtained do not depend on the bar 
potential we use, but on the presence of homoclinic or heteroclinic orbits.
Thus, if the model does not have either heteroclinic nor homoclinic orbits and
only transit orbits are present, the barred galaxy will present two spiral arms
emanating from the ends of the bar. The outer branches of the unstable invariant 
manifolds will spiral out from the ends of the bar and they will not return to
its vicinity. If the transit orbits associated to the $W^u_{\gamma_1}$ intersect
in configuration space with the transit orbits associated to $W^u_{\gamma_2}$, then
they form the characteristic shape of $R_2$ rings. That is, the trajectories 
outline an outer ring whose principal axis is parallel to the bar major axis. If 
heteroclinic orbits exist, then the ring of the galaxy is classified as $rR_1$.
The inner branches of the invariant manifolds associated to $\gamma_1$ and
$\gamma_2$ outline a nearly circular inner ring that encircles the bar. The outer
branches of the same invariant manifolds form an outer ring whose principal
axis is perpendicular to the bar major axis. The last possibility is if only
homoclinic orbits exist. In this case, the inner branches of the invariant
manifolds for an inner ring, while the outer branches outline both types of
outer rings, thus the barred galaxy presents an $R_1R_2$ ring morphology.

\section{Discussion}
\label{sec:dis}

\begin{figure}[h]  
\centering 
\hspace{1.cm}
\includegraphics[scale=0.3,angle=-90.]{romerogomez_png_fig4.ps}\hspace{0.55cm}
\includegraphics[scale=0.3,angle=-90.]{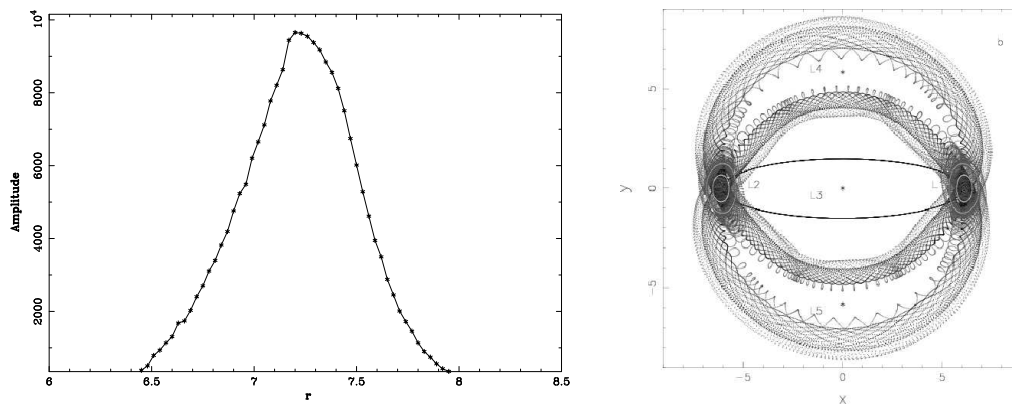}
\caption{{\it Left panel:} Density profile on a cut across the ring.
{\it Right panel:} Two unstable invariant manifolds for different
values of $E_J$. Note how similar the regions they delineate are.} 
\label{myfig3} 
\end{figure}

The family of Lyapunov orbits is unstable and becomes stable only at high
energy levels (Skokos {\it et al.} (\cite{sko02}). We compute the
invariant manifolds of Lyapunov orbits in the range of energies in which
they are unstable. In the right panel of Fig. \ref{myfig3}, we plot
the invariant manifolds for two different energy levels of a given model.
We find that the locus of the invariant manifolds is independent of the
energy. As the energy increases, however, the size of the Lyapunov orbits
also increases and thus the size of the invariant manifolds. Nevertheless,
as we consider more energy levels, we find that the density in the central
part is higher (see left panel of Fig. \ref{myfig3}). Therefore, the
thickness of the ring observed is smaller than the thickness of the
invariant manifold of higher energies. We also compute the radial and 
tangential velocities on the galactic plane and in a non-rotating reference 
frame along the ring and we observe that they are small perturbations of the 
circular velocity. The maximum deviation from the typical circular velocity 
of $200 \rm{km}\,\rm{s}^{-1}$ is $\pm 20 \rm{km}\,\rm{s}^{-1}$ (Athanassoula
{\it et al.} \cite{ath07}).

In the case of the spiral arms, we compute the density profile on two different 
angles, namely one near the beginning of the arm and one at the end. We find
that the first cut has a narrow and high density profile, while the second cut, 
has a wide and low density profile. This is the typical behaviour of the grand 
design spiral arms, namely they are denser and brighter near the bar ends and 
then they become more diffuse (Athanassoula {\it et al.} \cite{ath07}). 

To summarise, our results show that invariant manifolds describe well the loci 
of the different types of rings and spiral arms. They are formed by a bundle of
trajectories linked to the unstable regions around the $L_1/L_2$ equilibrium
points. The study of the influence of one model parameter on the shape of
the invariant manifolds in the outer parts reveals that only the pattern
speed and the bar strength affect the galaxy morphology. The study also shows
that the different ring types and spirals are obtained when we vary the
model parameters.

We have compared our results with some observational data. Regarding the
photometry, the density profiles across radial cuts in rings and spiral
arms agree with the ones obtained from observations. The velocities along
the ring also show that these are only a small perturbation of the circular
velocity.

\begin{acknowledgements} I wish to thank my collaborators E. Athanassoula, 
J.J. Masdemont, and C. Garc\'ia-G\'omez.
\end{acknowledgements}

\end{document}